\documentclass[twocolumn,twocolappendix,trackchanges]{aastex6}
\RequirePackage{lineno}
\usepackage{amsmath}
\usepackage{amssymb}
\usepackage{amsthm}
\usepackage{natbib,aasdefs,url,bm}
\usepackage{array}
\usepackage{float}
\usepackage{graphicx}
\usepackage{subfigure}
\usepackage{color}

\newcounter{ichi}
\newcounter{ni}
\newcounter{san}
\newcounter{yon}
\def\be{\begin{equation}}
\def\ee{\end{equation}}
\def\ba{\begin{eqnarray}}
\def\ea{\end{eqnarray}}

\newcommand{\ergsec}{\mbox{erg~s$^{-1}$}}

\newcommand{\gevcmsqs}{\mbox{GeV~cm$^{-2}$~s$^{-1}$}}

\def\aap{\ {A\&A}\ }

\def\apjl{\ {ApJL}\ }
\def\apjs{\ {ApJS}\ }

\shorttitle{Multimessenger Constraints on Neutrino Production in NGC 1068}
\shortauthors{Murase}

\begin{document}

\title{Hidden Hearts of Neutrino Active Galaxies}

\author{Kohta Murase\altaffilmark{1,2,3}}

\altaffiltext{1}{Department of Physics; Department of Astronomy \&
Astrophysics; Center for Multimessenger Astrophysics, Institute for Gravitation and the Cosmos, The Pennsylvania State University, University Park, PA 16802, USA}
\altaffiltext{2}{School of Natural Sciences, Institute for Advanced Study, Princeton, NJ 08540, USA}
\altaffiltext{3}{Center for Gravitational Physics and Quantum Information, Yukawa Institute for Theoretical Physics, Kyoto University, Kyoto, Kyoto 606-8502, Japan}

\begin{abstract}
Recent multimessenger studies have provided evidence for high-energy neutrino sources that are opaque to GeV--TeV gamma rays. We present model-independent studies on the connection between neutrinos and gamma rays in the active galaxy NGC 1068, and find that the neutrinos most likely come from regions within $\sim30-100$ Schwarzschild radii. This is especially the case if neutrinos are produced via the photomeson production process, although the constraints could be alleviated if hadronuclear interactions are dominant.   
We consider the most favorable neutrino production regions, and discuss coronae, jets, winds, and their interactions with dense material. The results strengthen the importance of understanding dissipation mechanisms near the coronal region and the outflow base. There could be a connection between active galactic nuclei with near-Eddington accretion and tidal disruptions events, in that neutrinos are produced in the obscured vicinity of supermassive black holes.  
\end{abstract}

%\keywords{galaxies: active -- galaxies: jets -- neutrinos -- radiation mechanisms: non-thermal}

\section{Introduction}
The origin of high-energy cosmic neutrinos has been a big enigma in particle astrophysics since their discovery by the IceCube Collaboration~\citep{Aartsen:2013bka,Aartsen:2013jdh}. Among various candidate sources considered in the literature~\citep[see recent reviews, e.g.,][]{Halzen:2022pez,Kurahashi:2022utm}, high-energy neutrino sources opaque to high-energy gamma rays, or ``hidden'' cosmic-ray (CR) accelerators, have been of interest. They have been required by the multimessenger connection between the all-sky neutrino flux in the 10~TeV range and the diffuse isotropic gamma-ray background \citep{Murase:2015xka}. Identifying the hidden neutrino sources enables us to utilize neutrinos as a unique probe of dense environments that cannot be studied only with electromagnetic observations. 

Recently, the IceCube Collaboration reported an excess of 79 events associated with a nearby spiral galaxy known as M77 or NGC 1068~\citep{IceCube2022NGC1068}. The reported significance is $4.2\sigma$, which was found in a search defined {\it a priori}, using a list of the catalog of sources observed at gamma rays and/or other wavelengths, and the result strengthens the previous report of a $2.9\sigma$ excess~\citep{IceCube:2019cia}. NGC 1068 is known to be a prototypical Seyfert II galaxy, which is a type of active galactic nucleus (AGN), as well as one of the starburst galaxies. A supermassive black hole (SMBH) at the center and its surroundings are highly obscured by thick gas and dust~\citep[e.g.,][]{GarciaBurillo+16ALMA,GamezRosas+22}, while X-ray studies have suggested that NGC 1068 is among the brightest AGNs in intrinsic X-rays~\citep{Bauer:2014rla,Marinucci:2015fqo,Ricci:2017dhj}. NGC 1068 has been considered as a promising neutrino source in light of both AGN~\citep{Murase:2019vdl,Kheirandish:2021wkm,Anchordoqui:2021vms,Inoue:2022yak} and starburst~\citep{Yoast-Hull:2013qfa,Murase:2016gly,Lamastra:2016axo} activities.  

In this work, we apply model-independent multimessenger analysis used in \cite{Murase:2013rfa} and \cite{Murase:2015xka} to single source emission. The approach is general and different from the other previous model-dependent studies on neutrino emission from NGC 1068. In Section~\ref{sec:general}, for the first time, we obtain constraints on the neutrino emission radius of NGC 1068, and show that neutrino production most likely occurs in the vicinity of the SMBH, especially near the coronal region. In Sections~\ref{sec:mechanism} and \ref{sec:acceleration}, based on the constraints obtained in Section~\ref{sec:general}, we discuss viable neutrino production mechanisms as well as specific models for particle acceleration. 
We use $Q_x=Q/10^{x}$ in cgs units and assume cosmological parameters with $\Omega_m=0.3$, $\Omega_\Lambda=0.7$, and $h=0.7$.

%%%%%%%%%%%%%%%%%%%%%%%%%%%%%%%%%%
\begin{figure}[t]
\includegraphics[width=\linewidth]{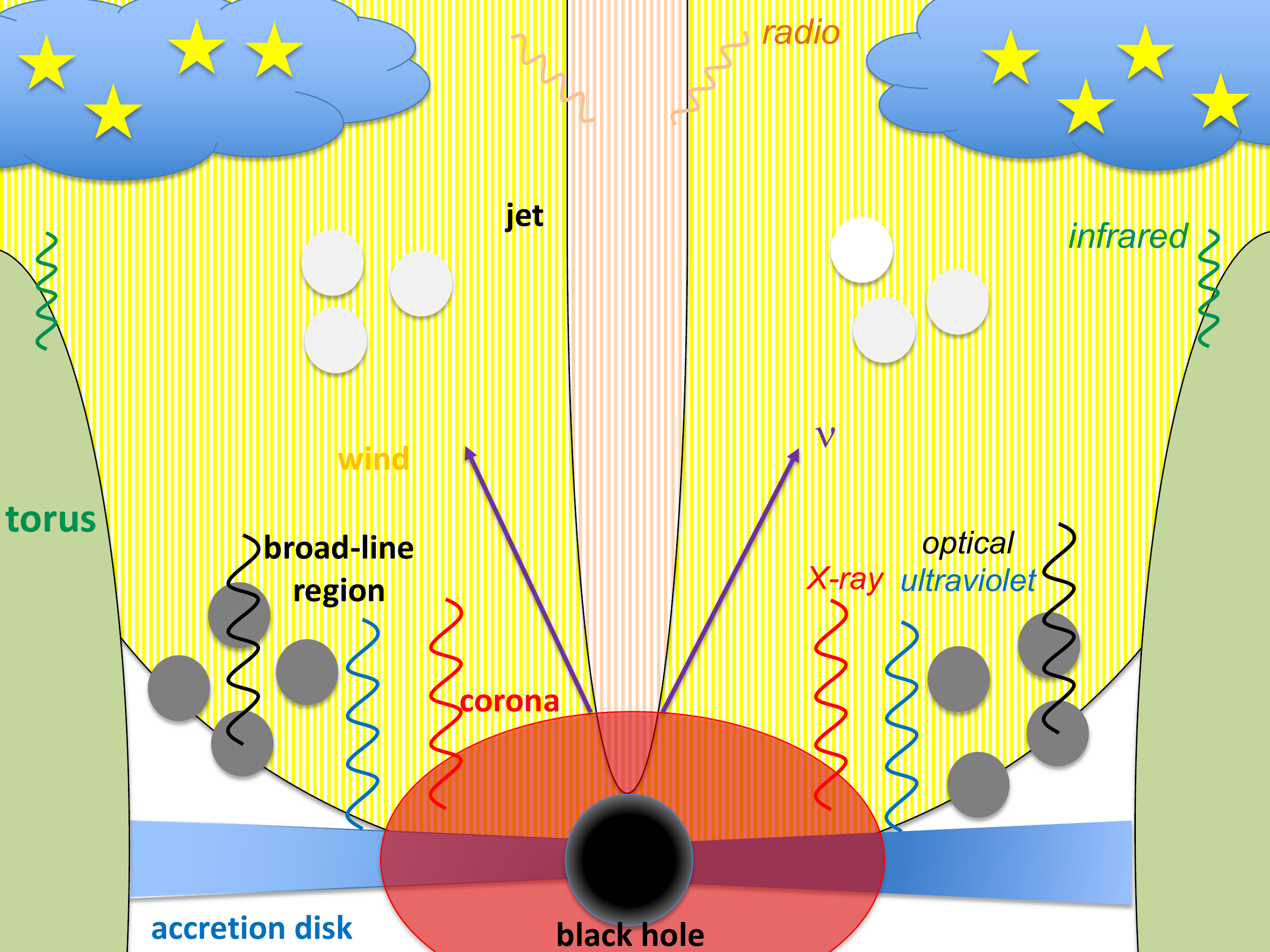}
\caption{Schematic picture of an AGN that produces high-energy neutrinos. Gas accreting onto an SMBH forms an accretion disk and hot corona, from which optical, ultraviolet, and X-rays are emitted. Winds and jets may also be launched. Infrared radiation comes from a dusty torus and the starburst region. Electromagnetic emission from the disk, corona, and broad-line regions is highly obscured. 
\label{fig:AGN}
}
%\vspace{-1.\baselineskip}
\end{figure}
%%%%%%%%%%%%%%%%%%%%%%%%%%%%%%%%%%

\section{Neutrino--Gamma-Ray Connection and Model-independent Constraints}
\label{sec:general}
High-energy neutrinos are produced through meson production by $p\gamma$ and/or $pp$ interactions. In either case, neutrino emission must be accompanied by gamma-ray production, and the differential luminosities of generated neutrinos (for all flavors) and pionic gamma rays (from $\pi^0\rightarrow2\gamma$) are approximately related as~\citep[see Equation~(3) of][]{Murase:2015xka}
\begin{equation}
\varepsilon_\gamma L_{\varepsilon_\gamma}^{(\rm gen)}\approx \frac{4}{3K}{[\varepsilon_\nu L_{\varepsilon_\nu}]}_{\varepsilon_\nu=\varepsilon_\gamma/2},
\label{eq:mm}
\end{equation}
where $K=1$ ($K=2$) for $p\gamma$ ($pp$) interactions, and $\varepsilon$ is particle energy in the source frame. The neutrino flux observed on Earth is given by $E_\nu F_{E_\nu}=\varepsilon_\nu L_{\varepsilon_\nu}/(4\pi d_L^2)$, where $E=\varepsilon/(1+z)$ is particle energy on Earth and $d_L$ is the luminosity distance. Equation~(\ref{eq:mm}) suggests that the neutrino and gamma-ray fluxes are comparable.    

NGC 1068 is known to be a gamma-ray source, and the gamma-ray flux measured by {\it Fermi} LAT is $E_\gamma F_{E_\gamma}\sim10^{-9}$~\gevcmsqs\ in the $0.1-100$~GeV range~\citep{Ackermann:2012vca,Fermi-LAT:2019yla}. On the other hand, MAGIC placed upper limits, $E_\gamma F_{E_\gamma}\lesssim10^{-10}-10^{-9}$~\gevcmsqs\ at sub-TeV energies~\citep{MAGIC:2019fvw}. The all-flavor neutrino flux reported by IceCube is $E_\nu F_{E_\nu}\sim10^{-7}$~\gevcmsqs\ around 1~TeV~\citep{IceCube:2019cia,IceCube2022NGC1068}, which is significantly higher than the {\it Fermi} gamma-ray flux and upper limits in the TeV range. In this sense, NGC 1068 has to be a hidden neutrino source. 
Indeed, in the simple starburst galaxy model, hadronic emission that is calibrated by the {\it Fermi} data is difficult to explain the IceCube flux~\citep[see Figure 4 of][]{Murase:2019vdl}. 

What are the implications of this opaqueness? The emission radius is one of the important quantities in modeling of high-energy source emission. In this section, we show that the neutrino emission radius can now be constrained thanks to the new IceCube data~\citep{IceCube2022NGC1068} as well as the existing multiwavelength observations in infrared, optical, X-ray, and gamma-ray bands.

\subsection{Attenuation Argument}
High-energy gamma rays from AGNs interact with photons from the accretion disk and hot corona, line emission from broad-line regions (BLRs), and infrared emission from the dusty torus (see Figure~\ref{fig:AGN}). 
The SMBH mass of NGC 1068 is estimated to be $M\sim(1-2)\times{10}^7~M_\odot$~\citep{Woo:2002un,Panessa+06} and the Schwarzschild radius is given by $R_S\equiv 2 GM/c^2\simeq5.9\times{10}^{12}~{\rm cm}~M_{7.3}$, where $M=10^{7.3}~M_{7.3}~M_\odot$. Within $\sim10^4~R_{\rm S}$ corresponding to the typical BLR radius at $R_{\rm BLR}\approx{10}^{17}~{\rm cm}~L_{\rm disk,45}^{1/2}$, where $L_{\rm disk}$ is the accretion disk luminosity, the most important radiation fields are disk and corona components. For the two-photon annihilation process, $\gamma\gamma\rightarrow e^+e^-$, the typical energy of a photon interacting with a gamma ray is $\tilde{\varepsilon}_{\gamma\gamma-\gamma}= m_e^2c^4{\varepsilon_\gamma}^{-1}\simeq0.26~{\rm keV}~(1~{\rm GeV}/\varepsilon_\gamma)$. 

\begin{figure}[t]
\includegraphics[width=\linewidth]{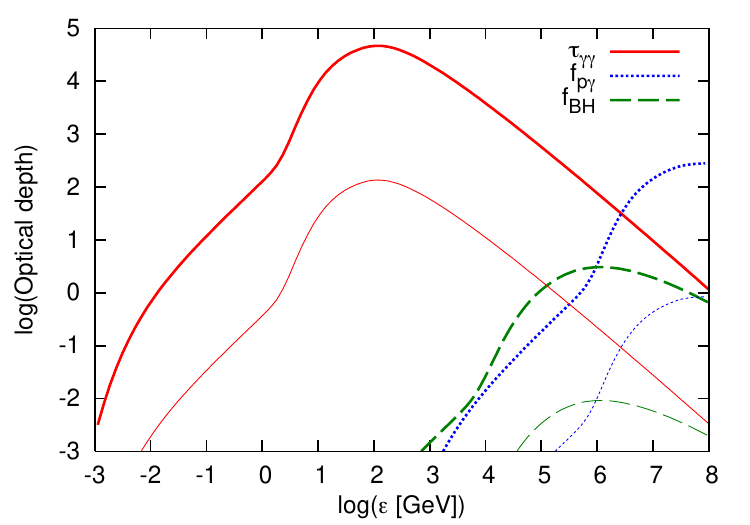}
\caption{Optical depths for two-photon pair annihilation, photomeson production and Bethe-Heitler pair production processes. The thick and thin lines are for $R=30~R_S$ and $R={10}^4~R_S$, respectively, and $t_*=10R/c$ is assumed.   
\label{fig:optdep}
}
%\vspace{-1.\baselineskip}
\end{figure}

In Figure~\ref{fig:optdep}, we present numerical results of the optical depth to $\gamma\gamma\rightarrow e^+e^-$, $\tau_{\gamma\gamma}(\varepsilon_\gamma)$, for different values of the emission radius $R\equiv{\mathcal R}R_S$, where ${\mathcal R}$ is the dimensionless emission radius. 
For the disk component, we assume a multitemperature blackbody spectrum expected for a standard disk
%~\citep{Shakura:1972te} 
with a bolometric luminosity of $L_{\rm bol}=10^{45}~{\rm erg}~{\rm s}^{-1}$~\citep[e.g.,][]{Woo:2002un,Zaino:2020elj} and the maximum energy, $\varepsilon_{\rm disk}=31.5$~eV~\citep{Inoue:2022yak}. For the corona component, we use the results of {\it NuSTAR} and {\it XMM-Newton} observations~\citep{Marinucci:2015fqo,Bauer:2014rla}, which suggest that the intrinsic X-ray luminosity (before the attenuation) is $L_X=7_{-4}^{+7}\times10^{43}~{\rm erg}~{\rm s}^{-1}$ (in the $2-10$~keV band) and a photon index of $\Gamma_X\approx2$ with a possible cutoff energy of $\varepsilon_{X,\rm cut}=128$~keV.  

As seen from Figure~\ref{fig:optdep}, the optical depth for GeV--TeV gamma rays is quite large due to the disk component, and their escape from the source is difficult even at $R=10^{4}~R_S\sim R_{\rm BLR}$. This also suggests that multi-GeV or lower-energy gamma rays can escape for emission radii beyond the BLR radius. For $R=30~R_S$, which is comparable to the size of the corona, even GeV gamma rays do not escape, and the source can be transparent to $\sim10$~MeV or lower-energy gamma rays.  

In general, high-energy gamma rays do not have to be observed as they are because they interact with ambient photons, and initiate electromagnetic cascades.  
Equation~(\ref{eq:mm}) has been employed as precascaded spectra to compare the IceCube neutrino data to the gamma-ray data particularly in the context of intergalactic cascades~\citep[][]{Murase:2013rfa,Murase:2015xka,Capanema:2020rjj,Capanema:2020oet,Fang:2022trf}. In this work, we apply this intrinsic multimessenger connection to {\it intrasource cascades}. If a cascade is fully developed via inverse-Compton (IC) emission and two-photon annihilation, the resulting spectrum is approximated by~\citep[e.g.,][]{Murase:2012df,Fang:2022trf}
%Berezinsky:1975zz,
\begin{equation}
\varepsilon_\gamma L_{\varepsilon_\gamma}^{(\rm cas)}
%\sim f_{\rm cas}\frac{Y_{\rm IC}}{1+Y_{\rm IC}}E_\gamma F_{E_\gamma}^{\rm gen}
\propto
\left\{
\begin{array}{lr} 
{(\varepsilon_{\gamma}/\varepsilon_{\gamma}^b)}^{1/2} & \mbox{($\varepsilon_\gamma < \varepsilon_{\gamma}^{b}$)}  \,\,\, \,\,\, \,\,\, \,\,\, \,\,\, \,\,\,\\
{(\varepsilon_{\gamma}/\varepsilon_{\gamma}^{b})}^{2-s_{\rm cas}} & \mbox{($\varepsilon_{\gamma}^{b} \leq \varepsilon_\gamma < \varepsilon_\gamma^{\rm cut}$)} 
\end{array} 
\right., 
\label{eq:cas}
\end{equation}
where $s_{\rm cas}\sim2$, $\varepsilon_\gamma^b\approx(4/3){(\varepsilon_\gamma^{\rm cut}/2[m_ec^2])}^2\varepsilon_{\rm disk}\simeq0.038~{\rm GeV}~{(\varepsilon_\gamma^{\rm cut}/1~{\rm GeV})}^2(\varepsilon_{\rm disk}/30~\rm eV)$ and $\varepsilon_\gamma^{\rm cut}$ is the gamma-ray energy at $\tau_{\gamma\gamma}(\varepsilon_\gamma)=1$. Although the normalization of the cascade flux depends on details, $E_\gamma F_{E_\gamma}\sim (0.1-0.5)E_\nu F_{E_{\nu}}$ is typically expected within 1 order of magnitude. Thus, for the cascaded flux not to violate the {\it Fermi} data, the source has to be opaque to gamma rays in the $0.1-10$~GeV range. These gamma rays mainly interact with X-rays from the corona, and the two-photon annihilation optical depth for $\Gamma_X\approx2$ is
\begin{eqnarray}
\tau_{\gamma\gamma}(\varepsilon_\gamma)&\approx&
\eta_{\gamma\gamma}\sigma_{T}R n_X{(\varepsilon_\gamma/\tilde{\varepsilon}_{\gamma\gamma-X})}^{\Gamma_X-1}\nonumber\\
&\simeq&99~{\mathcal R}_{1.5}^{-1}M_{7.3}^{-1}f_{\Omega}^{-1}\tilde{L}_{X,43.64}(\varepsilon_\gamma/1~\rm GeV),
\label{eq:gg}
\end{eqnarray}
where $\eta_{\gamma\gamma}\sim0.1$ is a coefficient depending on the target photon spectrum~\citep{sve87}, $\sigma_T\approx6.65\times{10}^{-25}~{\rm cm}^2$, $\tilde{\varepsilon}_{\gamma\gamma-X}=m_e^2c^4/\varepsilon_X\simeq0.26~{\rm GeV}~{(\varepsilon_X/1~{\rm keV})}^{-1}$, $\varepsilon_X$ is the reference X-ray energy, and $n_X\approx \tilde{L}_X/(4\pi f_{\Omega} R^2 c \varepsilon_X)$ is used. Here $\tilde{L}_X$ is the differential X-ray luminosity and $f_{\Omega}\equiv\Delta\Omega/(4\pi)$ is a geometrical factor. 
Requiring $\tau_{\gamma\gamma}(\varepsilon_\gamma=0.3~{\rm GeV})\gtrsim 10$ for gamma rays not to escape, we obtain ${\mathcal R}\equiv R/R_S\lesssim100$ as a constraint on the size of the neutrino production region, which suggests that neutrinos from NGC 1068 are produced in the vicinity of the SMBH.

%%%%%%%%%%%%%%%%%%%%%%%%%%%%%%%%%%
%%%%%%%%%%%%%%%%%%%%%%%%%%%%%%%%%%

\subsection{Cascade Constraints}
The estimates provided in the previous subsection may be too rough. 
In reality, not only gamma rays from $\pi^0$ decay but also electrons and positrons from $\pi^\pm$ decay and the Bethe-Heitler pair production contribute to the gamma-ray flux observed on Earth. Here we present results of more sophisticated analysis through the calculation of electromagnetic cascades inside the source. We numerically calculate cascade spectra following \cite{Murase:2017pfe} and \cite{Murase:2019vdl} with detailed secondary spectra of $pp$ and $p\gamma$ interactions~\citep{Kelner:2006tc,Kelner:2008ke}.
NGC 1068 is known to be a Compton thick AGN with $N_H\gtrsim1.5\times10^{24}~{\rm cm}^{-2}$. With a column density of $N_H=10^{25}~{\rm cm}^{-2}$, we also take into account gamma-ray attenuation due to the Compton scattering and pair production process,
%~\citep{Murase:2014bfa}, 
although our results are not much affected by such attenuation in matter. (For example, the pair production optical depth in matter is $\sim\alpha_{\rm em}\sigma_T N_H\lesssim1$, where $\alpha_{\rm em}=1/137$ is the fine structure constant.) TeV or higher-energy photons are unlikely to escape due to $\gamma\gamma\rightarrow e^+e^-$ with infrared emission from the dusty torus, and we implement this as attenuation using a blackbody spectrum with $T_{\rm dust}=1000$~K~\citep{Inoue:2022yak}. 
Finally, we impose $E_\gamma F_{E_\gamma}^{(\rm cas)}<E_\gamma F_{E_\gamma}^{(\rm obs)}$ to evaluate the constraints on $\mathcal R$.  

%%%%%%%%%%%%%%%%%%%%%%%%%%%%%%%%%%
\begin{figure*}[t]
\begin{center}
\includegraphics[width=0.32\linewidth]{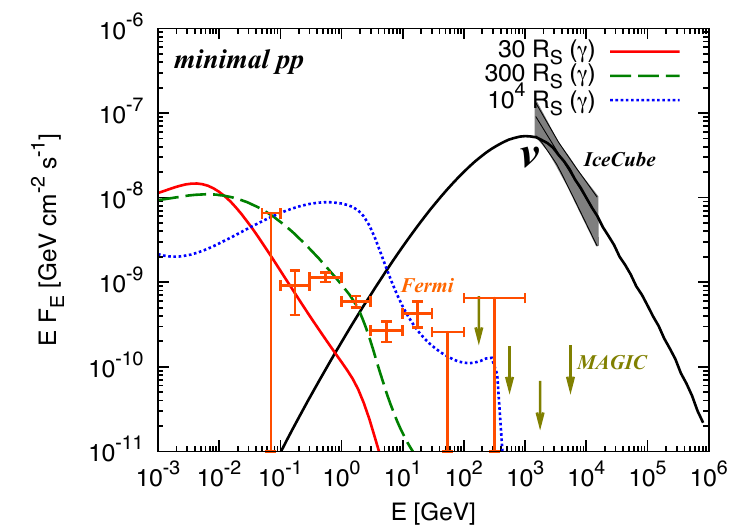}
\includegraphics[width=0.32\linewidth]{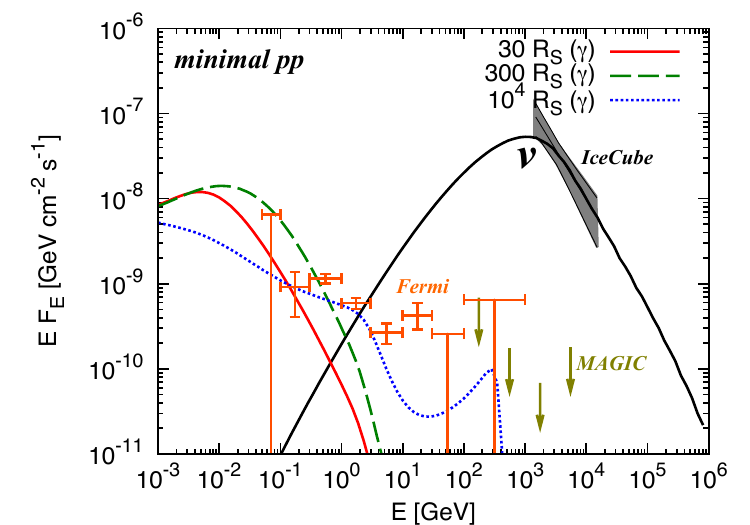}
\includegraphics[width=0.32\linewidth]{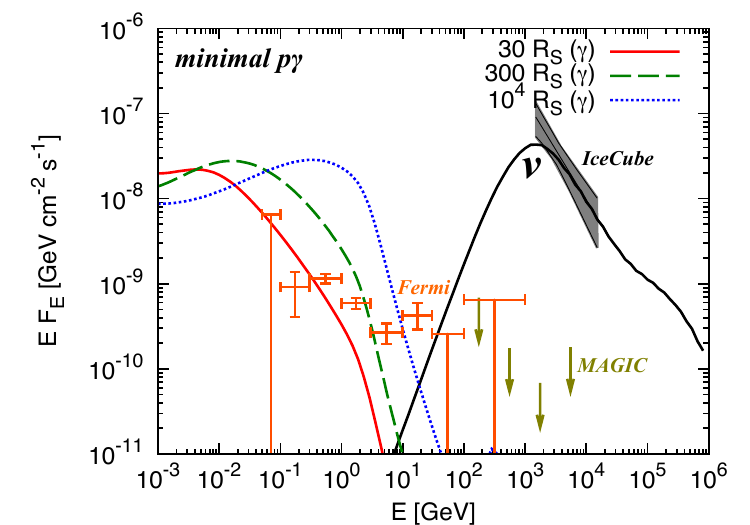}
\caption{
Left: neutrino and cascaded gamma-ray spectra in the minimal $pp$ scenario with $\xi_B=0.01$, where the IC cascade contribution is significant. 
Middle: same as the left panel but for $\xi_B=1$, where the synchrotron cascade dominates. 
Right: neutrino and cascaded gamma-ray spectra in the minimal $p\gamma$ scenario with $\xi_B=1$, where the Bethe-Heitler pair production enhances the cascade flux.
}
\label{fig:cascade}
%\vspace{-1.\baselineskip}
\end{center}
\end{figure*}
%%%%%%%%%%%%%%%%%%%%%%%%%%%%%%%%%%

Our cascade constraints rely on the IceCube data and radiation fields inferred by electromagnetic observations, independent of details of intrinsic CR spectra. This is because, as shown in Equation~(\ref{eq:mm}), the relationship between neutrinos and gamma rays is uniquely determined by particle physics, whether neutrinos are produced by $pp$ interactions (hadronuclear scenarios) or $p\gamma$ interactions (photohadronic scenarios). Our approach is different from the other previous studies that calculate CR spectra under model assumptions. The minimum scenarios described below have only two generic parameters, which are the emission radius and the magnetic field.

We parameterize the magnetic field by $\xi_B\equiv U_B/U_\gamma$, where $U_B=B^2/(8\pi)$ and $U_\gamma=L_{\rm bol}/(4\pi f_{\Omega}R^2c)$, and we have $B\simeq 1.4\times{10}^3~{\rm G}~\xi_B^{1/2}f_{\Omega}^{-1/2}L_{\rm bol,45}^{1/2}M_{7.3}^{-1}{\mathcal R}_{1.5}^{-1}$. 
%We adopt $f_{\Omega}=1$ for demonstration.  
The values of $\xi_B$ are model dependent, and we demonstrate cases of $\xi_{B}=0.01$ and $\xi_{B}=1$. For outflow models, $\xi_{B}\lesssim0.01$ are typically expected. For example, in the wind model, the magnetic field is written as $U_B=\epsilon_B L_{w}/(4\pi f_{w}R^2 V_w)$ with the equipartition parameter $\epsilon_B$, where observations of supernova remnants indicate $\epsilon_B\sim{10}^{-3}-{10}^{-2}$~\citep[e.g.,][]{Vink:2011ei}, and $L_{w}=(1/2)(\eta_w/\eta_{\rm rad})L_{\rm bol}{(V_w/c)}^2\simeq5.6\times{10}^{43}~\ergsec~(\eta_w/\eta_{\rm rad,-1})L_{\rm bol,45}V_{w,9.5}^2$ is the wind luminosity, $f_w$ is the wind solid angle, $V_w\sim(0.003-0.3)$~c is the wind velocity (that depends on the launching radius), $\eta_w$ is the wind efficiency, and $\eta_{\rm rad}\sim0.1$ is the radiative efficiency. 
%We have $B\simeq 1.0\times{10}^2~{\rm G}~\epsilon_{B,-2}^{1/2}f_{w}^{-1/2}{(\eta_w/\eta_{\rm rad,-1})}^{1/2}L_{\rm bol,45}^{1/2}M_{7.3}^{-1}{\mathcal R}_{1.5}^{-1}V_{w,9.5}^{1/2}$. 
We have $\xi_B\simeq0.53\times{10}^{-2}~\epsilon_{B,-2}{(f_\Omega/f_{w})}{(\eta_w/\eta_{\rm rad,-1})}V_{w,9.5}$. 
NGC 1068 has a weak jet with a luminosity of $L_{j}\sim{\rm a~few}\times 10^{43}~{\rm erg}~{\rm s}^{-1}$~\citep{GarciaBurillo+14}. In the relativistic jet model as another example, we have $U_B=\epsilon_B L_j/(4\pi f_j R^2\Gamma_j^2c)\approx \epsilon_B L_j/(2\pi R^2c)$ and 
%$B\simeq20~{\rm G}~\epsilon_{B,-2}^{1/2}L_{j,43}^{1/2}M_{7.3}^{-1}{\mathcal R}_{1.5}^{-1}$.  
$\xi_B\simeq0.63\times{10}^{-2}~\epsilon_{B,-1}f_\Omega (L_{j,43.5}/L_{\rm bol,45})$, where the beaming factor is $f_j\approx1/{(2\Gamma_j^2)}$, $\Gamma_j$ is the jet Lorentz factor, and $\epsilon_B\sim{10}^{-4}-{10}^{-1}$ from the literature of gamma-ray burst afterglows~\citep[e.g.,][]{GvdH14}.
More generally, $\xi_B\sim1$ is possible, e.g., in the magnetically powered corona model~\citep{Murase:2019vdl}, where the magnetic field can be estimated by $B={(8\pi n_p^{\rm cor} kT_p^{\rm cor}/\beta)}^{1/2}$, leading to $\xi_B\simeq1.4~\beta^{-1}{(\tau_T^{\rm cor}/\zeta_e)}f_{\Omega}M_{7.3}L_{\rm bol,45}^{-1}$. 
%$\simeq 1.7\times{10}^3~{\rm G}~\beta^{-1/2}{(\tau_T^{\rm cor}/\zeta_e)}^{1/2}M_{7.3}^{-1/2}{\mathcal R}_{1.5}^{-1}$~\citep{Murase:2019vdl}, 
Here $\beta$ is the plasma beta, $n_{p}^{\rm cor}$ is the coronal proton density, $kT_p^{\rm cor}$ is the proton temperature that is set to the virial temperature, $\tau_{T}^{\rm cor}\sim0.1-1$ is the coronal optical depth, and $\zeta_e$ is the pair loading factor. 

First, we consider {\it hadronuclear ($pp$) scenarios}. These scenarios are commonly considered for gamma-ray transparent neutrino sources~\citep{Murase:2013rfa}. For an $E_\nu^{-2}$ spectrum, i.e., $E_\nu F_{E_\nu}\propto const.$, the constraint from Equation~(\ref{eq:gg}), ${\mathcal R}\lesssim100$, should still be applied not to violate the {\it Fermi} data. 
More generally, harder CR spectra are possible if CRs are accelerated via magnetic reconnections and/or stochastic acceleration. Thus we consider the minimum $pp$ scenario, in which the spectrum has a low-energy cutoff at $\varepsilon_p=10$~TeV to explain the neutrino spectrum only above $1.5$~TeV (see Figure~\ref{fig:cascade} left and middle). This can mimic models where the CR spectrum is effectively harder than $dL_{\rm CR}/d\ln {\varepsilon}_p\propto {\varepsilon}_p^{1.3}-{\varepsilon}_p^{1.5}$~\citep{Murase:2019vdl}. Note that due to the energy distribution of $pp$ yields and pion/muon decay, the neutrino spectrum cannot have an abrupt cutoff even in such a minimal scenario~\citep{Murase:2015xka}. 

Our numerical results of the minimal $pp$ scenario for $\xi_B=0.01$ are shown in Figure~\ref{fig:cascade} (left). IceCube data of neutrinos~\citep{IceCube2022NGC1068}, {\it Fermi} data of gamma rays~\citep{Fermi-LAT:2019yla}, and MAGIC gamma-ray upper limits~\citep{MAGIC:2019fvw} are also depicted. In this case, the IC cascade is important, and at 100~GeV the IC cooling time of electrons and positrons is comparable to their synchrotron cooling time. The emission radius is constrained to be ${\mathcal R}\lesssim100$ for $\xi_B\lesssim0.01$, consistent with the previous analytical constraint.

The results for $\xi_B=1$ are shown in Figure~\ref{fig:cascade} (middle). Synchrotron emission by secondary pairs contributes to sub-GeV emission for small values of $\mathcal R$, so the range of $100\lesssim {\mathcal R}\lesssim3000$ is disfavored. When gamma rays are suppressed by the two-photon annihilation process, we again obtain ${\mathcal R}\lesssim100$, consistent with the previous estimates.
On the other hand, the range of ${\mathcal R}\sim(3\times{10}^3-{10}^5)$ is allowed because of the synchrotron cascade. While such parameter space is possible, the following conditions need to be met: (i) the CR spectrum is very hard and nearly monoenergetic; (ii) the magnetization is stronger than those expected in outflow models but the magnetic field should not be too strong for the synchrotron cascade flux to overshoot the {\it Fermi} data. 
In the corona model, such a strong magnetization is plausible but ${\mathcal R}\gtrsim3\times{10}^3$ is unlikely. 

Finally, the results for {\it photohadronic ($p\gamma$) scenarios} with $\xi_B=1$ are shown in Figure~\ref{fig:cascade} (right). As found in \cite{Murase:2019vdl}, the Bethe-Heitler pair production process plays a crucial role in the presence of disk-corona radiation fields, which enhances the IC cascade flux compared to the case only with photomeson production. The Klein-Nishina suppression is less important for the Bethe-Heitler pairs because for a given $\varepsilon_p$ they have $\sim100$ times lower energies than the pairs from pion/muon decay. We obtain $\mathcal R\lesssim30$ for $\xi_B\lesssim1$.

\subsection{Energetics and Meson Production Efficiency}
The differential neutrino luminosity around 1~TeV is $\varepsilon_\nu L_{\varepsilon_\nu}\sim3\times{10}^{42}$~\ergsec~\citep{IceCube2022NGC1068}, so the inferred CR luminosity is 
\begin{eqnarray}
L_{\rm CR}&\approx&\frac{4(1+K)}{3K}{\rm max}[1,f_{\rm mes}^{-1}]f_{\rm sup}^{-1} (\varepsilon_\nu L_{\varepsilon_\nu}) {\mathcal C}_{p}\nonumber\\
&\sim&8\times{10}^{42}~{\ergsec}~{\rm max}[1,f_{\rm mes}^{-1}]f_{\rm sup}^{-1}{\mathcal C}_{p},
\end{eqnarray}
where $K=1$ is used, $f_{\rm mes}$ is the effective meson production optical depth (see below), $f_{\rm sup}$ is the suppression factor e.g., due to the CR Bethe-Heitler cooling, and ${\mathcal C}_{p}$ is the correction factor from the differential luminosity to the bolometric luminosity.  
Sufficiently hard CR spectra below $\varepsilon_p\sim10$~TeV are required from the energetics. If the CR spectrum is extended to GeV energies with a spectral index of $s_{\rm CR}=3.2$, we have $C_p\gtrsim6\times{10}^4$, which violates the upper limit from the bolometric luminosity. Even with $s_{\rm CR}=2.0$, we have $C_p\gtrsim9$, leading to $L_{\rm CR}\sim L_X \sim 0.1 L_{\rm bol}$. The constraint on $L_{\rm CR}$ becomes more severe for $f_{\rm mes}f_{\rm sup}<1$, and the calorimetric condition, $f_{\rm mes}\geq 1$, is favored. 

The meson production cannot be inefficient, and $L_{\rm CR}\lesssim L_{\rm bol}$ leads to $f_{\rm mes}\gtrsim0.06-0.08$, which is the most conservative constraint (although the bolometric luminosity is uncertain). In $p\gamma$ scenarios, another constraint can be obtained by the opacity argument \citep[e.g.,][]{Murase:2015xka}. The photomeson production optical depth can be written as 
\begin{eqnarray}
f_{p\gamma}(\varepsilon_p^c)&\approx&\frac{\eta_{p\gamma}\hat{\sigma}_{p\gamma}}{\eta_{\gamma\gamma}\sigma_{\gamma\gamma}}\tau_{\gamma\gamma}(\varepsilon_\gamma)\left(\frac{t_*}{R/c}\right)\nonumber\\
&\sim&0.1~\left(\frac{\tau_{\gamma\gamma}(\varepsilon_\gamma)}{10}\right)\left(\frac{t_*}{10R/c}\right)\,,
\label{eq:opticaldepth}
\end{eqnarray}
where $\hat{\sigma}_{p\gamma}\sim0.7\times{10}^{-28}~{\rm cm}^2$ and $\eta_{p\gamma}\sim1$ is a coefficient depending on the target photon spectrum and multipion production, $\varepsilon_p^{c}$ is the proton energy corresponding to the gamma-ray energy satisfying $\varepsilon_p^c\approx [m_p\bar{\varepsilon}_\Delta/(2m_e^2c^2)]\varepsilon_\gamma \sim 160~{\rm TeV}~(\varepsilon_\gamma/0.3~{\rm GeV})$, and $\bar{\varepsilon}_\Delta\sim0.3$~GeV.
%Note that the takes into account the case where thh source is nonrelativistic~\citep[cf.][]{Murase:2015xka}. 
In a steady-state system, the characteristic time for CRs to get exposed to photons ($t_{*}$) is the CR escape time. Hereafter to discuss cases for the most efficient neutrino production we consider $t_{*}\approx R/V$ in the limit that the CRs are confined in the plasma, where $V$ is the characteristic velocity, e.g., the infall velocity $V_{\rm fall}$ in accretion flows and the shock velocity $V_s$ in shock models. See also Figure~\ref{fig:optdep}. Then, from $\tau_{\gamma\gamma}(\varepsilon_\gamma=0.3~{\rm GeV})\gtrsim 10$, we obtain $f_{p\gamma}(\varepsilon_p^c)\gtrsim0.1~(0.1c/V)$. 

\section{Neutrino Production Mechanisms}
\label{sec:mechanism}
It is natural to ask where the calorimetric condition, i.e., $f_{\rm mes}\geq1$, can be satisfied. 
In this section, we address this question and discuss viable scenarios for neutrino production in NGC 1068. All proposed models~\citep{Murase:2019vdl,Inoue:2019yfs,Kheirandish:2021wkm,Anchordoqui:2021vms,Eichmann:2022lxh,Inoue:2022yak} can be understood as specific cases allowed by the constraints provided in Section~\ref{sec:general}.

\subsection{Hidden Photohadronic Scenarios}
A large fraction of Seyfert galaxies and quasars are believed to have a geometrically thin, radiation-dominated disk with a typical energy of $\varepsilon_{\rm disk}\sim 10-30$~eV. It is known that $\sim10$\% of the bolometric luminosity is radiated as X-rays that presumably originate from hot plasma regions called coronae. These two components are regarded as almost guaranteed target photon fields (although in NGC 1068 their intrinsic fluxes are subject to significant uncertainties due to the opaqueness).   
We do not consider radiation from electrons that are coaccelerated with protons because details of the dissipation and particle acceleration are uncertain.

\subsubsection{Direct Emission from Accretion Disks and Coronae}
In the previous section, we found that the neutrino emission radius in $p\gamma$ scenarios is $R\lesssim30~R_S$, for which direct emission from the disk and corona are the most important. Neutrinos with $\varepsilon_\nu\sim1~{\rm TeV}$ are mostly produced by $\sim20$~TeV protons interacting with X-rays, so coronal X-rays provide target photons for the photomeson production, and for $\Gamma_X\approx2$ we have~\citep{Murase:2019vdl} 
\begin{eqnarray}
f_{p \gamma}
%&\approx&t_{\rm esc}/t_{p\gamma}\nonumber\\
&\approx&\eta_{p\gamma}\hat{\sigma}_{p\gamma}R(c/V) n_X{(\varepsilon_p/\tilde{\varepsilon}_{p\gamma-X})}^{\Gamma_X-1}\nonumber\\
&\sim&0.9~\frac{\eta_{p\gamma}\tilde{L}_{X,43.64}(V_{\rm fall}/V)}{\alpha_{-1}{\mathcal R}_{1.5}^{1/2}M_{7.3}}{\left(\frac{\varepsilon_p}{30~\rm TeV}\right)},
%&\sim&4.32\frac{\eta_{p\gamma}L_{X,43.64}(V_{\rm fall}/V){(\varepsilon_p/\tilde{\varepsilon}_{p\gamma-X})}^{\Gamma_X-1}}{\alpha_{-1}{\mathcal R}_{1.5}^{1/2}M_{7.3}(\varepsilon_X/1~{\rm keV})},
\label{eq:pg}
\end{eqnarray}
where $n_X=\tilde{L}_X/(2\pi R^2 c \varepsilon_X)$ is used given that $\tau_T^{\rm cor}\lesssim1$, and $\tilde{\varepsilon}_{p\gamma-X}=0.5m_pc^2\bar{\varepsilon}_{\Delta}/\varepsilon_X\simeq0.14~{\rm PeV}~{(\varepsilon_X/1~{\rm keV})}^{-1}$. 
Here, according to the corona model, the infall velocity is used as a fiducial value, $V_{\rm fall}\approx \alpha V_K\simeq 1.3\times{10}^{-2}c~\alpha_{-1}{\mathcal R}_{1.5}^{-1/2}$, where $V_K$ is the Keplerian velocity and $\alpha\sim0.1$ is the viscosity parameter. 
From Equation~(\ref{eq:pg}), we expect that $f_{p\gamma}\gtrsim1$ for protons making neutrinos with $E_\nu\approx1.5$~TeV is satisfied at ${\mathcal R}\lesssim30$. 
In the accretion shock model~\citep[e.g.,][]{Inoue:2019fil}, the freefall velocity $V_{\rm ff}\approx V_K\simeq 0.13c~{\mathcal R}_{1.5}^{-1/2}$ is used so that $f_{p\gamma}$ is lower and the calorimetric condition is not satisfied even at ${\mathcal R}\sim1$.
\cite{Inoue:2022yak} considered failed winds with $V=10^{8}~{\rm cm}~{\rm s}^{-1}$, in which the IceCube flux is explained by $p\gamma$ neutrinos. 
For $V\gtrsim {10}^{-3}c$, the calorimetric condition for TeV neutrinos leads to ${\mathcal R}\lesssim100$.    

Note that photomeson production due to disk photons can be important at higher energies, because the threshold energy is $\tilde{\varepsilon}_{p\gamma-\rm th}\simeq1.1~{\rm PeV}~{(\varepsilon_{\rm disk}/30~{\rm eV})}^{-1}$. Jet-quiet AGNs have been suggested as the sources of PeV neutrinos~\citep{Stecker:1991vm,Kalashev:2015cma}, although such a scenario requires fast CR acceleration mechanisms such as shock acceleration.    

As noted above, CR protons responsible for $\sim1-100$~TeV neutrinos mainly lose energies via the Bethe-Heitler process (see Figure~\ref{fig:optdep}).  %because the characteristic energy is $\tilde{\varepsilon}_{\rm BH-disk}\approx0.5m_pc^2\bar{\varepsilon}_{\rm BH}/\varepsilon_{\rm disk}\simeq0.47~{\rm PeV}~{(\varepsilon_{\rm disk}/10~{\rm eV})}^{-1}$, where $\bar{\varepsilon}_{\rm BH}\sim10(2m_ec^2)\sim10$~MeV~\citep{1992ApJ...400..181C,Murase:2010va}. 
With the disk photon density $n_{\rm disk}\sim L_{\rm disk}/(2\pi R^2 c \varepsilon_{\rm disk})$, the effective Bethe-Heitler optical depth (with $\hat{\sigma}_{\rm BH}\sim0.8\times{10}^{-30}~{\rm cm}^2$) is
\begin{eqnarray}
f_{\rm BH}&\approx&n_{\rm disk}\hat{\sigma}_{\rm BH}R(c/V)\nonumber\\
&\sim&40~\frac{L_{\rm disk,45}(V_{\rm fall}/V)}{\alpha_{-1}{\mathcal R}_{1.5}^{1/2}M_{7.3}{(\varepsilon_{\rm disk}/30~{\rm eV})}},\,\,\,\,\, 
%&\sim&37.8~L_{\rm disk,45}\alpha_{-1}^{-1}{\mathcal R}_{1.5}^{-1/2}M_{7.3}^{-1}{(30~{\rm eV}/\varepsilon_{\rm disk})},\,\,\,\,\, 
\end{eqnarray}
as the maximum value around $\varepsilon_p\sim1$~PeV (see Figure~\ref{fig:optdep}), which implies $f_{\rm sup}\approx f_{p\gamma}/f_{\rm BH}\sim0.1-1$ in $p\gamma$ scenarios~\citep{Murase:2019vdl}. This can make the energetics constraint somewhat more severe. 

With the calorimetric limit of ${\rm min}[1,f_{p\gamma}]=1$, the all-flavor neutrino flux is
\begin{eqnarray}
E_\nu F_{E_\nu}&\approx&\frac{1}{8\pi d_L^2} \left[\frac{3K}{2(1+K)}\right]f_{\rm sup} \frac{L_{\rm CR}}{{\mathcal C}_{p}}\nonumber\\ 
&\sim&1\times{10}^{-7}~\gevcmsqs\nonumber\\
&\times&~\left[\frac{3K}{2(1+K)}\right]f_{\rm sup,-0.5}L_{\rm CR,43.5}{\mathcal C}_p^{-1},\,\,\,\,\,\,\,\,\,\,\,
\label{eq:nuflux}
\end{eqnarray}
which can be compatible with the observed neutrino flux~\citep{IceCube:2019cia,IceCube2022NGC1068}. 

Target photons from the disk and corona are mainly produced at the vicinity of the SMBH, $\lesssim10-100~R_{S}$. If the emission region is much larger, their photon density decreases as $R^{-2}$ and the efficiency is reduced. 
%If the emission region relativistically moves outward, the photon energy is reduced by $\Gamma$, changing the threshold condition

\subsubsection{Scattered and Reprocessed Emission}
A significant fraction of disk-corona photons are scattered or reprocessed by gas and dust inside the BLR and dusty torus~\citep{Netzer:2015jna,RamosAlmeida:2017bbn}. The energy density of these photons including broad-line emission is
\begin{equation}
U_\gamma \approx \frac{\tau_{\rm sc}^{\rm eff}(L_{\rm disk}+L_{X})+L_{\rm BL}}{4 \pi R_{\rm BLR}^2},
\end{equation}
where $\tau_{\rm sc}^{\rm eff}\lesssim f_{\rm cov}$ is the effective scattering optical depth, $L_{\rm BL}\approx f_{\rm cov}L_{\rm disk}$ is the luminosity of broad-line emission, and $f_{\rm cov}$ is the covering factor of the BLR clouds, which is typically $\sim0.1$ but may be larger for Seyfert II galaxies~\citep{RamosAlmeida:2017bbn}.
%EsparzaArredondo+21

The density of these scattered and reprocessed photons is lower than that of direct emission from the disk and corona at $R<R_{\rm BLR}$. However, because these components are nearly isotropic in the SMBH rest frame, they are more important for high-energy emission from jetted AGNs including blazars~\citep{Dermer:2014vaa}.

\subsection{Hidden Hadronuclear Scenarios}
Hadronuclear ($pp$) scenarios are commonly considered in the context of gamma-ray transparent sources~\citep{Murase:2013rfa}. Hidden $pp$ scenarios are also possible~\citep[e.g.,][]{Kimura:2014jba}, 
%Kimura:2019yjo
but target material is different among models.   

\subsubsection{Accretion Flows}
CRs may interact with accreting gas via inelastic $pp$ collisions. 
The plasma density in the accretion flows of a geometrically thin and radiation-dominated disk is very high but particle acceleration in such collisional plasma would not be efficient. Although CRs accelerated at different regions could hit the disk, we do not consider such a possibility. A more promising site for CR acceleration may be the coronal region, and neutrinos can be produced via $pp$ interactions with the coronal plasma. The coronal proton density is $n_{\rm cor}\approx\tau_{T}^{\rm cor}/(\zeta_e\sigma_T H)$, where $H\approx R/\sqrt{3}$ is the scale height. The effective $pp$ optical depth is~\citep{Murase:2022feu} 
\begin{eqnarray}
f_{pp}
%&\approx&t_{*}/t_{pp}
&\approx& n_p^{\rm cor} \hat{\sigma}_{pp}R(c/V)\nonumber\\
&\sim&2~(\tau_T^{\rm cor}/0.4)\zeta_e^{-1}\alpha_{-1}^{-1}{\mathcal R}_{1.5}^{1/2}(V_{\rm fall}/V),
%&\simeq&0.98~(\tau_T^{\rm cor}/0.4)\zeta_e^{-1}\alpha_{-1}^{-1}{\mathcal R}_{1.5}^{1/2}(V_{\rm fall}/V),
\end{eqnarray}
where $\hat{\sigma}_{pp}\sim2\times{10}^{-26}~{\rm cm}^2$ is the attenuation $pp$ cross section including proton inelasticity. Given $\tau_T^{\rm cor}\sim0.1-1$, inelastic $pp$ interactions are efficient in the coronal region if the pair loading is not significant~\citep{Murase:2019vdl,Eichmann:2022lxh}. 
Recent magnetohydrodynamics (MHD) simulations have suggested that such a coronal region exists at $\mathcal R\lesssim5-50$~\citep[e.g.,][]{Jiang:2014wga,JBSS19,Chashkina:2021zox,Liska:2022jdy}, but it is unclear whether the calorimetric condition can be satisfied at larger radii. 

It has been shown that $pp$ neutrinos from the coronal region~\citep{Murase:2019vdl,Inoue:2019yfs} can explain the observed neutrino flux of NGC 1068\footnote{The neutrino flux of \cite{Murase:2019vdl} shown in \cite{IceCube2022NGC1068} corresponds to $L_X\sim(3-4)\times{10}^{43}$~\ergsec~for $d=12.7-14.4$~Mpc. The corona model has uncertainty coming from the intrinsic X-ray luminosity~\citep{Kheirandish:2021wkm}.} if the CR luminosity is $\sim10-100\%$ of the X-ray luminosity ($\sim1-10\%$ of the bolometric luminosity). 
The all-flavor neutrino flux in $pp$ scenarios can also be estimated by Equation~(\ref{eq:nuflux}), but with $f_{\rm sup}\approx f_{pp}/f_{\rm BH}$ if $f_{\rm BH}\gtrsim1$.

\subsubsection{Interacting Outflows}
Outflows (jets and winds) can be driven by the accretion disk and/or the spinning SMBH. CRs may be accelerated by these outflows via shocks and/or magnetic reconnections, and $pp$ interactions may occur inside the outflows. For the winds, the mass-loss rate is $\dot{M}_w=\eta_wL_{\rm bol}/(\eta_{\rm rad}c^2)$, leading to the proton density, $n_p^w\simeq4.8\times{10}^9~{\rm cm}^{-3}~f_{w}^{-1}{(\eta_w/\eta_{\rm rad,-1})}L_{\rm bol,45}M_{7.3}^{-2}{\mathcal R}_{1.5}^{-2}V_{w,9.5}^{-1}$, and we have $f_{pp}\approx n_p^w\hat{\sigma}_{pp}R(c/V_w)\simeq0.17~f_{w}^{-1}{(\eta_w/\eta_{\rm rad,-1})}L_{\rm bol,45}M_{7.3}^{-1}{\mathcal R}_{1.5}^{-1}V_{w,9.5}^{-2}$. The observational indication of weak jets with $L_{j}\sim{\rm a~few}\times 10^{43}~{\rm erg}~{\rm s}^{-1}$ allows us to consider CR acceleration in jets near the SMBH. Although the jet may be magnetically dominated and its kinetic energy may be carried by pairs, even with a comoving proton density of ${n'}_p^j\simeq4.0\times{10}^7~{\rm cm}^{-3}~{(1+\sigma_{\rm mag})}^{-1}L_{j,43.5}M_{7.3}^{-2}{\mathcal R}_{1.5}^{-2}$ (where $\sigma_{\rm mag}$ is the magnetization parameter), the effective $pp$ optical depth is $f_{pp}\approx{n'}_p^j\hat{\sigma}_{pp}(R/\Gamma_j)\ll 1$. Thus, neutrino production inside the outflows would not be efficient enough. 

More efficient $pp$ interactions may happen when these outflows interact with dense material. 
%The wind shock $\sim0.01-10$~kpc~\citep{Liu:2017bjr}, at which gamma rays can escape so this model is unlikely to explain the neutrino flux of NGC 1068.  
%Jet-cloud interaction model is hard for explaining blazar neutrinos, it may be possible for AGNs like NGC 1068 given that neutrino emission is not beamed. 
NGC 1068 shows a large column density of $N_H>10^{24}~{\rm cm}^{-2}$~\citep[e.g.,][]{Marinucci:2015fqo,Zaino:2020elj}. Models have suggested that multiple absorbers and reflectors are necessary, and they are presumably located from the BLR radius to the inner radius of the dusty torus with $R_{\rm DTin}\sim0.1-0.2$~pc, where the wind-torus interaction is indicated~\citep{GarciaBurillo+16ALMA,GarciaBurillo+19windtorus}. 
The effective $pp$ optical depth can be~\citep{Inoue:2022yak}
\begin{eqnarray}
f_{pp}&\approx&n_{\rm wout}\hat{\sigma}_{pp}c t_{*}\nonumber\\
&\sim&0.6~n_{\rm wout,7}R_{\rm DTin,17.5}V_{w,9.5}^{-1}\left(\frac{t_*}{R_{\rm DTin}/V_{w}}\right),\,\,\,\,\,\,\,\,\,\,
\end{eqnarray}
where $n_{\rm wout}$ is the density of the wind-torus interface.

BLR clouds may work as X-ray absorbers %~\citep[e.g.,][]{2002ApJ...571..234R,2014MNRAS.439.1403M}, 
and the dusty torus may also be clumpy, which would be responsible for changes of the X-ray flux~\citep[e.g.,][]{Marinucci:2015fqo,Zaino:2020elj}.  
The formation mechanism of these clouds is still uncertain but it has been suggested that they may form in disk-driven winds~\citep[e.g.,][for a review]{Netzer:2015jna}. The density of BLR clouds, $n_{\rm cl}$, is around $10^9-10^{11}\rm~cm^{-3}$ and the cloud size, $R_{\rm cl}$, is a few $R_S$. Then the column density per cloud can be $n_{\rm cl}R_{\rm cl}=10^{24}~{\rm cm}^2~n_{\rm cl,11}R_{\rm cl,13}$.
%Note that details depend on the number of clouds, which can be $N_{\rm cl}\sim 10^7-10^8$ clouds in an AGN \citep{1998MNRAS.297..990A,1999AA...351...31D}.
If the jet or wind interacts with a dense cloud, a bow shock with velocity $\sim V_w$ and a slower cloud shock with $V_{\rm cl}$ are formed. From the pressure balance, $n_{\rm cl}V_{\rm cl}^2\approx n_p^{w}V_w^2$, the forward shock crosses the cloud with $V_{\rm cl}\simeq3.9\times{10}^{7}~{\rm cm}~{\rm s}^{-1}~n_{\rm cl,11}^{-1/2}f_{w}^{-1/2}{(\eta_w/\eta_{\rm rad,-1})}^{1/2}L_{\rm bol,45}^{1/2}M_{7.3}^{-1}{\mathcal R}_{1.5}^{-1}V_{w,9.5}^{1/2}$, and the cloud may eventually evaporate.  
%the total mass of BLRs is $10^4-10^5$ $\rm M_{\odot}$. This mass is consistent with the estimate by the CLOUDY model \citep{2003ApJ...582..590B}, although it is much higher than the standard expectation of $1-10\rm~M_{\odot}$.
The effective $pp$ optical depth is estimated to be
\begin{eqnarray}
f_{pp}\approx n_{\rm cl}\hat{\sigma}_{pp}c t_{*}
&\simeq&15~n_{\rm cl,11}^{3/2}R_{\rm cl,13}f_{w}^{1/2}{(\eta_w/\eta_{\rm rad,-1})}^{-1/2}\nonumber\\
&\times&L_{\rm bol,45}^{-1/2}
M_{7.3}{\mathcal R}_{1.5}V_{w,9.5}^{-1/2}
\left(\frac{t_*}{R_{\rm cl}/V_{\rm cl}}\right).\,\,\,\,\,\,\,\,\,\,
\end{eqnarray}
In principle, efficient neutrino production could occur. However, the overall efficiency should be reduced by the covering factor $f_{\rm cov}$ because only CRs encountering clouds can produce neutrinos. Also, only sufficiently high-energy CRs would be able to enter the clouds, and a low-energy cutoff is expected. 

The neutrino flux may be written as
\begin{eqnarray}
E_\nu F_{E_\nu}&\approx&\frac{1}{8\pi d_L^2}{\rm min}[1,f_{pp}]f_{\rm cov}\frac{L_{\rm CR}}{{\mathcal C}_{p}}\nonumber\\ 
&\lesssim&1\times{10}^{-7}~\gevcmsqs~f_{\rm cov,-0.5} L_{\rm CR,43.5}{\mathcal C}_{p}^{-1}.\,\,\,\,\,\,\,\,\,\,
\end{eqnarray}
Because of the cascade constraints, extreme conditions (very large $\epsilon_B$ and small ${\mathcal C}_p$ for shock acceleration or very large $L_{\rm CR}$) seem to be required to achieve the IceCube flux without violating the {\it Fermi} data. As demonstrated with the wind-torus interaction model~\citep{Inoue:2022yak}, the {\it Fermi} gamma-ray flux could be explained, but the neutrino flux is typically lower than the IceCube data.

\section{Possible Particle Acceleration Sites}
\label{sec:acceleration}
The observed neutrino spectrum is soft with an index of $s_\nu=3.2$~\citep{IceCube2022NGC1068}, for which there are two possible explanations. 
The first possibility is that the soft spectrum is actually the tail of a hard CR spectrum, which is the case in many models~\citep{Murase:2019vdl,Inoue:2019yfs,Eichmann:2022lxh,Inoue:2022yak}. For example, for a CR spectral index of $s_{\rm CR}=2$, the neutrino spectrum has to have a high-energy cutoff or break at $E_\nu\sim10$~TeV~\citep{Kheirandish:2021wkm}, which corresponds to the proton maximum energy, $\varepsilon_p^{\rm max}\sim200$~TeV.
The second possibility is that the spectrum of accelerated CRs is intrinsically soft. However, as noted in Section~\ref{sec:general}, such a spectrum cannot extend down to GeV energies not to exceed the bolometric luminosity and a low-energy break or cutoff is necessary. 
If the system is calorimetric for meson production without $f_{\rm sup}$, the CR spectral index above $\sim10$~TeV is also $s_{\rm CR}\sim3.2$, but the inferred index is model dependent. 
In the presence of $f_{\rm sup}<1$ due to the Bethe-Heitler process, the injected CR spectrum is harder. 
On the other hand, $p\gamma$ scenarios with $f_{\rm p\gamma}, f_{\rm BH}\lesssim1$ may imply a soft CR spectrum with $s_{\rm CR}\sim s_\nu+1=4.2$. 

Although particle acceleration sites are highly uncertain and further studies are necessary, we discuss several possibilities in this section.   

\subsection{Coronae}
%
%%%%%%%%%%%%%%%%%%%%%%%%%%%%%%%%%%
\begin{figure}[t]
\includegraphics[width=\linewidth]{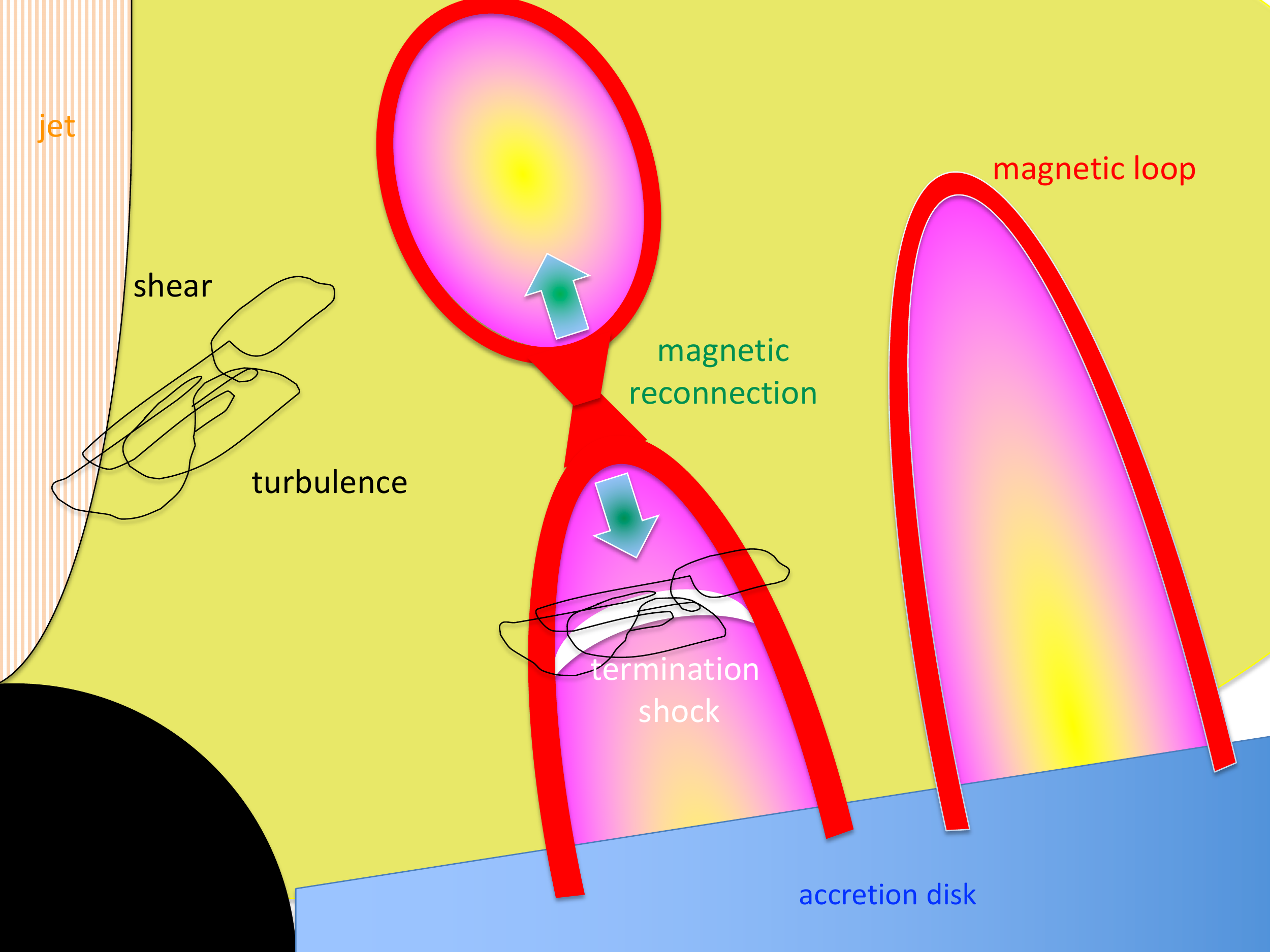}
\caption{Schematic picture of possible particle acceleration sites in the coronal regions near an SMBH. 
\label{fig:corona}
}
%\vspace{-1.\baselineskip}
\end{figure}
%%%%%%%%%%%%%%%%%%%%%%%%%%%%%%%%%%

Neutrino emission radii of $R\lesssim30-100~R_S$ tempt us to consider particle acceleration near the coronal region. The existence of coronae consisting of hot low-$\beta$ plasma has been indicated in MHD simulations~\citep[e.g.,][]{Jiang:2014wga,JBSS19,Chashkina:2021zox,Liska:2022jdy}, and CR acceleration via magnetic dissipation is invoked for neutrino emission~\citep[][]{Kheirandish:2021wkm}.  
%Not only large-scale magnetic fields from the magnetic buoyancy but also turbulent magnetic fields due to magnetorotational instability and magnetic reconnections. Details of CR acceleration are uncertain. 
Magnetic reconnections are likely to be important for the injection of high-energy particles~\citep{Khiali:2015tfa,Comisso:2022iqy}, and sufficiently high-energy CRs may be accelerated by MHD turbulence and shear~\citep[see Figure~\ref{fig:corona};][]{Kimura:2016fjx,Kimura:2018clk}. 
The stochastic acceleration time scale is~\citep{Murase:2019vdl}
\begin{eqnarray}
t_{\rm acc}&=&\eta_{\rm tur}{\left(\frac{c}{V_A}\right)}^2\frac{H}{c}{\left(\frac{\varepsilon_p}{eBH}\right)}^{2-q}\nonumber\\
&\simeq&3.6\times{10}^4~{\rm s}~\eta_{\rm tur,1}\beta{\mathcal R}_{1.5}^2M_{7.3} {\left(\frac{\varepsilon_p}{eBH}\right)}^{2-q},
\label{eq:stoacc}
\end{eqnarray}
where $\eta_{\rm tur}^{-1}$ is the energy fraction of turbulence, $q$ is the energy dependence of the momentum diffusion coefficient, and $V_A\simeq0.1c~\beta^{-1/2}{\mathcal R}_{1.5}^{-1/2}$ is the Alfv\'en velocity. 
Notably, for $q\sim2$ that corresponds to the hard sphere limit, $t_{\rm acc}$ is not far from the Bethe-Heitler cooling time scale, $t_{\rm BH}\approx4.7\times{10}^{4}~{\rm s}~L_{\rm disk,45}M_{7.2}^{-2}{\mathcal R}_{1.5}^{-2} (30~{\rm eV}/\varepsilon_{\rm disk})$ at 100~TeV. 
The acceleration time in the second-order Fermi mechanism is slower than the first-order one, and $t_{\rm acc}=t_{\rm BH}$ may lead to a maximum proton energy around $\varepsilon_p^{\rm max}\sim{\mathcal O}(100)$~TeV, consistent with the IceCube observation~\citep{Murase:2019vdl}. 

As the second possibility, we suggest a scenario that is analogous to the standard solar flare model. In solar flares~\citep[][for a review]{SM11}, the magnetic reconnection region is identified as a site of particle acceleration~\citep{Masuda+94}. A magnetic reconnection event leads to an upflow and a downflow. The latter forms a termination shock, at which CRs are accelerated~\citep{Chen:2015rcq}, although the postshock temperature is reduced via thermal conduction via magnetic loops~\citep{SM11}. 
In the AGN disk-corona system considered in this work, CRs could similarly be accelerated at the termination shock (see Figure~\ref{fig:corona}), in which the acceleration time scale for the diffusive shock acceleration mechanism is 
\begin{eqnarray}
t_{\rm acc}&=&\eta_{\rm sho}{\left(\frac{c}{V_A}\right)}^2\frac{\varepsilon_p}{eBc} \nonumber\\
&\simeq&11~{\rm s}~\eta_{\rm sho,1}B_{3}^{-1}{\left(\frac{0.1c}{V_A}\right)}^{2}{\left(\frac{\varepsilon_p}{100~{\rm TeV}}\right)},
\end{eqnarray}
where $V_s\sim V_A$ is used and $\eta_{\rm sho}$ is a coefficient that is known to be $20/3$ for a parallel shock in the Bohm limit. We have $\varepsilon_p^{\rm max}\sim$~a few~PeV for $\eta_{\rm sho}\sim10$, in which the CR spectrum has to be soft above 10~TeV, otherwise larger values of $\eta_{\rm sho}$ are necessary.
%required not to violate constraints on $\varepsilon_p^{\rm max}$. 

The jet may be launched in the polar region of the SMBH, presumably via the Blandford-Znajek mechanism. If the jet exists, high-energy CRs could also be accelerated by shear acceleration. Although details are beyond the scope of this work, its time scale is given by $t_{\rm acc}=\eta_{\rm she}\beta_{\Delta}^{-2}(l_{\rm tur}/c){[\varepsilon_p/(eBl_{\rm tur})]}^{2-q}\simeq6.2\times{10}^4~{\rm s}~\eta_{\rm she,1}(l_{\rm tur}/0.1R){\mathcal R}_{1.5}M_{7.3}\beta_{\Delta,-0.5}^{-2}{[\varepsilon_p/(eBl_{\rm tur})]}^{2-q}$, where $\eta_{\rm she}$ is an order-of-unity factor for shear acceleration, $\beta_{\Delta}\propto\varepsilon_p^{2-q}$ is the the velocity difference experienced by particles~\citep[e.g.,][]{Liu:2017gln} and $l_{\rm tur}$ is the characteristic turbulence scale that would be a sizable fraction of the system. The shear acceleration can be slow as in stochastic acceleration, and the maximum energy could be $\varepsilon_p^{\rm max}\sim{\mathcal O}(100)$~TeV for $q\sim2$ although details depend on the jet structure through $\beta_\Delta$.

\subsection{Shocks}
In the earliest models of AGN neutrinos, the accretion shock has been considered as a particle acceleration site~\citep[e.g.,][]{Stecker:1991vm}. In this scenario,  $V_s\approx V_{\rm ff}$ is expected, and $s_{\rm CR}\sim2$ has been assumed~\citep{Inoue:2019yfs,Anchordoqui:2021vms}, where $\eta_{\rm sho}$ has to be much larger than the canonical value in the Bohm limit not to violate the constraint on the neutrino break/cutoff~\citep[see Figure~5 of][]{Kheirandish:2021wkm}. 

Alternatively, shocks may be induced by collisions of fallback material~\citep{AlvarezMuniz:2004uz} 
%Ghisellini+04
or failed winds that are expected for line-driven winds from the accretion disk~\citep{Inoue:2022yak}.

%\begin{eqnarray}
%t_{\rm acc}&=&\eta_{\rm sho} \frac{\varepsilon}{eBc}{\left(\frac{c}{V_w}\right)}^2\nonumber\\
%&\simeq&11~{\rm s}~\eta_{\rm sho,1}B_{3}^{-1}{\left(\frac{0.1c}{V_w}\right)}^{2}{\left(\frac{\varepsilon_p}{100~{\rm TeV}}\right)}
%\end{eqnarray}

Successful disk-driven winds may also lead to shocks, at which CR protons can be accelerated~\citep[e.g.,][]{Lamastra:2016axo,Liu:2017bjr,Inoue:2022yak}. These shocks are considered at far regions of $R\gg {10}^4~R_S$, and protons may be accelerated to $\sim1-100$~PeV energies depending on the assumption for $\epsilon_B$ and $V_w$.

\section{Summary and Discussion}\label{sec:summary}
Not only the all-sky multimessenger fluxes but also the single source multimessenger spectra of NGC 1068 suggest hidden CR accelerators, in which GeV--TeV gamma rays are largely attenuated and cascaded inside the sources. By considering the neutrino and gamma-ray connection in NGC 1068, we showed that the neutrino production region most likely lies at $R\lesssim100~R_S$.  
This is particularly the case in $p\gamma$ scenarios due to the Bethe-Heitler pair production process, where we obtained $R\lesssim30~R_S$. We also found that outer-zone $pp$ scenarios, where GeV emission can be weak due to the synchrotron cascade, are possible although extreme conditions seem required. We stress that these new constraints, which can be placed thanks to the improved IceCube data~\citep{IceCube2022NGC1068}, are largely model independent and will be useful for the development of more sophisticated models. Also, the minimal scenarios give the most conservative constraints on $R$, and the limits can be stronger if the CR spectrum extends to lower energies and generate more gamma rays.

We considered where meson production occurs most efficiently (i.e., $f_{\rm mes}\gtrsim1$). 
Both $p\gamma$ and $pp$ interactions can be important in the vicinity of SMBHs. 
In hidden $p\gamma$ scenarios, ultraviolet and X-ray photons from the disk and corona are likely to play the major roles, and the calorimetric condition, $f_{p\gamma}\geq1$, suggests $R\lesssim100~R_S$, consistent with the cascade constraints. 
Alternatively, hidden $pp$ scenarios are possible but whether $f_{pp}\geq1$ is satisfied is model dependent, which could be realized in the corona or outflow-cloud interactions. 
%The calorimetric condition is more difficult satisfied if the charcteritstic velocity is the freefall velocity or faster. 

In summary, based on the general multimessenger constraints on $\mathcal R$ (Section~\ref{sec:general}) and calorimetric condition (Section~\ref{sec:mechanism}), the most favorable scenarios for hidden neutrino production in AGNs like NGC 1068 are   
\begin{itemize}
    \item $p\gamma$ scenarios at $R\lesssim30~R_S$, where CRs that are accelerated in coronae or by shocks interact with X-rays from the coronae. 
    \item $pp$ scenarios at $R\lesssim100~R_S$, where CRs that are accelerated in coronae or by shocks interact gas in inflowing material. 
\end{itemize}
We also provided $pp$ scenarios at $R\sim(3\times{10}^3-10^5)~R_S$, where CRs are accelerated and/or transported by outflows interacting with dense material, but they are less likely due to extreme conditions. This work assumed the persistence of neutrino emission, which is reasonable for AGNs although variability is naturally expected. One could consider a completely different type of a hidden neutrino source at far regions by assuming a neutrino luminosity of $\sim3\times{10}^{42}~{\rm erg}~{\rm s}^{-1}$ lasting for 10~yr. However, the required CR energy is $\gtrsim2\times{10}^{51}$~erg, which is rather extreme for transients like supernovae, and neutrinos from other galaxies would have been seen.

NGC 1068 was reported as the second compelling neutrino source~\citep{IceCube2022NGC1068}. The other candidate sources are blazar flares and tidal disruption events~\citep[TDEs; see reviews, e.g.,][]{Halzen:2022pez,Kurahashi:2022utm}, and none of them have reached the discovery level. Further data are needed, and because of the transient nature of blazars flares and TDEs as well as the lack of their concordance picture, NGC 1068 could become the first established source.

Remarkably, NGC 1068 is a near-Eddington accretion system with its Eddington parameter, $\lambda_{\rm Edd}\equiv L_{\rm bol}/L_{\rm Edd}\sim0.4$ (with significant uncertainty). Interestingly, all neutrino-associated TDEs are also near-Eddington accretion systems~\citep{vanVelzen:2021zsm}, although the Eddington ratios are lower at the time of neutrino detections. Both NGC 1068 and TDEs are hidden neutrino sources, where TDE neutrino production sites are also obscured by the debris and radiation fields by ultraviolet and X-ray emission. Perhaps high-energy neutrinos may be produced by the common mechanisms in AGNs and TDEs~\citep{Murase:2020lnu}. 

%For TXS 0506+056, the isotropic equivalent luminosity for neutrino flares is estimated to be $\sim 10^{46}-10^{47}~\ergsec$. This luminosity is larger than the Eddington luminosity of typical AGN of $M\sim10^8~M_{\odot}$. It is natural to consider that the neutrino emission region is relativistically beamed. $L_{\rm cr}\approx{10}^{48}~{\rm erg}~{\rm s}^{-1}~(8/K){\rm max}[1,f_{\rm mes}^{-1}]{\mathcal R}_{\rm cr}\theta_{j,-1}^2$, 

Our results imply that observations of high-energy neutrinos may give us new insight into plasma dissipation and particle acceleration in the vicinity of SMBHs. There has been significant progress in numerical simulations, and further theoretical studies will be useful for understanding the physics in obscured environments. 

Finally, because the neutrino flux is as large as $\sim10^{-7}$~\gevcmsqs\, we note that MeV and/or sub-GeV gamma-ray emission is expected as a consequence of electromagnetic cascades~\citep{Murase:2019vdl}, which are testable with future MeV gamma-ray detectors such as AMEGO-X~\citep{Caputo:2022xpx} and eASTROGAM~\citep{e-ASTROGAM:2016bph}. 

%%%%%%%%%%%%%%%%%%%%%%%%%%%%%%%%%%%%%%%%%%%%%%%%%%
%%%%%%%%%%%%%%%%%%%%%%%%%%%%%%%%%%%%%%%%%%%%%%%%%%

\begin{acknowledgements}
We thank Ali Kheirandish, Shigeo Kimura, Peter M\'esz\'aros and Shigeru Yoshida for comments or discussions. 
The work of K.M. is supported by the NSF grants No.~AST-1908689, No.~AST-2108466 and No.~AST-2108467, and KAKENHI No.~20H01901 and No.~20H05852.
\end{acknowledgements}

\bibliographystyle{aasjournal}
\bibliography{kmurase}

\end{document}